\documentclass[pdflatex,prb,twocolumn,showpacs]{revtex4}
\usepackage{graphicx}
\usepackage{dcolumn}
\usepackage{bm}
\usepackage{amsmath,epsf}
\usepackage[colorlinks, citecolor=red, breaklinks,pagebackref]{hyperref}

\begin{document}

\author{Dipta Bhanu Ghosh, and Stefano de Gironcoli}
\affiliation{DEMOCRITOS National Simulation Center, Trieste 34014, Italy \\
SISSA -- Scuola Internazionale Superiore di Studi Avanzati , Trieste 34014, Italy
}

\title{Structural and spin transitions in Fe$_{2}$O$_{3}$}

\date{\today}

\begin{abstract} 
First principles density functional calculations for Fe$_{2}$O$_{3}$
has been performed over a wide range of pressures. The ground state
is corundum-type hematite and is an antiferromagnetic insulator. This
is in good agreement with experiment and other theoretical studies. On
increasing pressure, the ground-state high-spin magnetic phase transforms
to low spin via the closure of the charge transfer gap. The system
also evolves to a new orthorhombic structure. Distorted corundum
or Rh$_{2}$O$_{3}$(II) type structure with Pbcn symmetry and Pbnm
type perovskite structure are two known competitive candidates for
this structural phase, based on {\it single}--cationic type and {\it
two}--cationic type picture, respectively. In our calculations, at about
38 GPa, Rh$_{2}$O$_{3}$(II) type structure becomes more stable with
respect to the ground state hematite. Relative stability of Pbnm type
perovskite is ruled out by our calculations in this pressure regime. The
Rh$_{2}$O$_{3}$(II) type structure remains in its low spin state, with 1
$\mu_{B}/$Fe atom, up to about 120 GPa. At this pressure the nonmagnetic
solution in Rh$_{2}$O$_{3}$(II) type structure becomes more favorable with
respect to the low spin one. By further increasing the pressure at about
330 GPa, the system evolves to yet another new structural phase. This new
orthorhombic structural phase is nonmagnetic and has Pmc2$_{1}$ symmetry,
a subgroup of Cmcm. Surprisingly, on furthering rising the pressure,
a Pbnm type nonmagnetic solution becomes competitive with the Pmc2$_{1}$
type structure and finally becomes stable at about 880 GPa.

\end{abstract}

\maketitle

\section{INTRODUCTION}

Iron (Fe), one of the most abundant element on earth, is believed to
contribute significantly to the mantle core characteristics in pure oxide
form (FeO, Fe$_{2}$O$_{3}$) or in mixed oxide forms such as (Mg,Fe)O and
(Mg,Fe)SiO$_{3}$. Of these oxides, Fe$_{2}$O$_{3}$ shows versatility
in it's structural, magnetic and electronic properties. The ambient
pressure phase $\alpha$-Fe$_{2}$O$_{3}$ (hematite) is corundum-type and
is a wide-gap antiferromagnetic insulator\cite{Hubbard64}. It becomes
weakly ferromagnetic between Morin temperature, T$_{M}$ (=260 K), and
N\'eel temperature, T$_{N}$ (= 955 K), as a result of the canting of
the spins of the two sublattices.

\par
Owing to its diverse intriguing properties it drew a significant amount
of attention in the scientific community from the sixties. Reid
and Ringwood\cite{Reid-jgr69} based on McQueen and Marsh's shock-wave
experiment\cite{Mcqueen-gsa66} proposed a new (denser) structural
phase to exist between 60 -- 120 GPa. In the subsequent years the
onset of an high pressure phase at $\sim$ 50 GPa was confirmed by
quite a few studies\cite{Shannon-jssc70, Yagi-capj82, Suzuki-ktksp85,
Olsen-ps91}. However, there was ambiguity concerning the crystal structure
of the new phase. While Shannon {\it et. al}\cite{Shannon-jssc70}
opted for the distorted corundum (Rh$_{2}$O$_{3}$-II) structure
speculating on a single cationic nature of Fe, {\it i.e.} Fe$^{3+}$,
the group of Suzuki {\it et. al.}\cite{Suzuki-ktksp85} went with the
idea of orthorhombic perovskite structure with ABO$_{3}$ formula with a
{\it two}--type of cationic Fe (Fe$^{2+}$ and Fe$^{4+}$) picture. Room
temperature M\"ossbauer spectroscopy (MS) experiments\cite{Suzuki-ktksp85,
Syono-ssc84, Nasu-hi86} revealed a non magnetic component at $\sim$ 50
GPa, coexisting with the magnetic phase. A volume collapse of $\sim$
10\%, concurrent with the onset of the phase transition at $\sim$ 50
GPa, corroborating a first order phase transition, was observed by Olsen
{\it et. al.}\cite{Olsen-ps91}. Assuming the two cationic type picture
this volume collapse was qualitatively explained by the increase in
coordination number (from 6 to 8) in one of the Fe sites.

\par
To resolve the issue of {\it one} or {\it two}--cationic types and
correlate/separate the spin and structural transitions a number of
investigations have been done in the more recent years. Of them, X-ray
diffraction (XRD), MS at 300K and electrical resistivity measurements
by Pasternak {\it et .al.}\cite{Pasternak-prl99} unequivocally assigned
high pressure structural phase as the nonmagnetic distorted corundum or
Rh$_{2}$O$_{3}$(II) type. They explained the intermediate region between
the insulating and metallic region in their electrical resistivity data
qualitatively by the coexistence of the insulating corundum phase (denoted
HP1 from now on) with the metallic Rh$_{2}$O$_{3}$(II) (denoted HP2 in
the following).

Other XRD and X-ray emission spectroscopy (XES) experiments by Badro {\it
et .al.}\cite{Badro-prl02} could separate the electronic transition from
the crystallographic transition. From their XES data it was concluded that
the high spin (HS) could be stabilized in the high pressure structural
phase, {\it i.e.} in HP2 phase, at low temperatures. At high temperatures,
however, the low spin (LS) state of HP2 is stable. That this LS state
is a weakly paramagnetic one was concluded by the presence of a satellite
(K$\beta^{\prime}$) in the XES spectra.\cite{Badro-prl02}

The room temperature Raman spectroscopy study by Shim and
Duffy\cite{Shim-am02} found the upper boundary for this phase transition 
(where no HP1 signal is left) to be between 54 -- 56 GPa on compression.
While decompressing, however, some (frequency) modes of the HP2 phase
survive down to 25 GPa. The combined experimental and theoretical
study by Rozenberg {\it et .al.}\cite{Rozenberg-prb02} also reported of
the sluggishness in the transition from HP1 to HP2. The onset of high
pressure phase start at about $\sim$ 40 GPa and at 76 GPa they could
find only the components of the high pressure phase. Their theoretical
calculations at 76 GPa predict the stability of the Rh$_{2}$O$_{3}$(II)
({\it i.e.} HP2) phase with respect to the orthorhombic perovskite phase
({\it opv} in the following).

The XRD study by Ono {\it et .al.}\cite{Ono-jpcs04,Ono-jpcm05} suggested
that high temperature heating has a major role to play in controlling
the phase transition pressure. The phase transition pressure of $\sim$
45 -- 50 GPa at room temperature drops down by about 15 -- 20 GPa when
the samples are heated to overcome the kinetic barriers hindering the
transition. Their estimated transition pressure was $\sim$ 26.2 GPa at
300K. However, in this study they were not conclusive about the crystal
structure of the high pressure phase.

There are a few theoretical studies
of the Fe$_{2}$O$_{3}$ system in the hematite
phase.\cite{Catti-prb95,Sandratskii-jpcm96,Punkkinen-jpcm99,Rollmann-prb04}
To the best of our knowledge, however, there are no theoretical
studies concerning the low and high pressures phases and their
stabilities. Neither the issue of transition from HS to LS is
addressed. In this work we will shed light on these issues of the
Fe$_{2}$O$_{3}$ system from first principles approach, based on density
functional theory (DFT).

\par
The paper is organized as follows: after a brief discussion of the
calculational details we will proceed to the next sections dealing
with the results of our calculations. We will start with the ambient
pressure phase, $\alpha$-Fe$_{2}$O$_{3}$ at the GGA level. After
that we will proceed with two (orthorhombic perovskite, Pbnm type and
Rh$_{2}$O$_{3}$(II), Pbcn type) competing candidate structures for the
high-pressure phase following the hematite one. The next subsection
will focus on the theoretical search for the experimentally found, but
structurally unresolved, post-perovskite type phase. 
At the end we will comment on the nature of the spin
transition and the evolution of the structure.

\section{CALCULATIONAL DETAILS}

We have performed first-principles density functional calculations to
analyze the spin and structural transitions in Fe$_2$O$_3$. Calculations
were performed using the PBE generalized gradient approximation
(GGA)\cite{Perdew-prl96}.  Variable-cell-shape molecular
dynamics\cite{Wentzcovitch-prb91} was used for the full structural
optimization at arbitrary pressures. Eight electrons (3d$^{7}$4s$^{1}$)
per Fe atom and six electrons (2s$^{2}$2p$^{4}$) per O atom were taken
in the valence in the generation of the ultrasoft pseudopotential
within the scheme of Rabe Rappe Kaxiras Joannopoulos ultrasoft
pseudopotential\cite{Rappe-prb90}. Total energy convergence of the wave
function cut-off and kinetic energy cut-off (for the charge density) were
checked. And the final values used were 35 and 380 Ryd, respectively. The
Brillouin zone integration was performed using a Monkhorst and Pack\cite{MP}
mesh
of 6\(\times\)6\(\times\)6 points. For the HP1 phase (corundum structure)
we used the rhombohedral cell consisting of 2 formula units ({\it f.u.})
of Fe$_{2}$O$_{3}$. For the HP2 phase, and possible HP3/HP4 phase,
orthorhombic cell consisting of 4 {\it f.u.} of Fe$_{2}$O$_{3}$ were used.

\section{RESULTS AND DISCUSSIONS}

\subsection{hematite}
\begin{figure}[tbp]
\begin{center}
\begin{minipage}[t]{0.36\linewidth}
\includegraphics[width=\linewidth,clip]{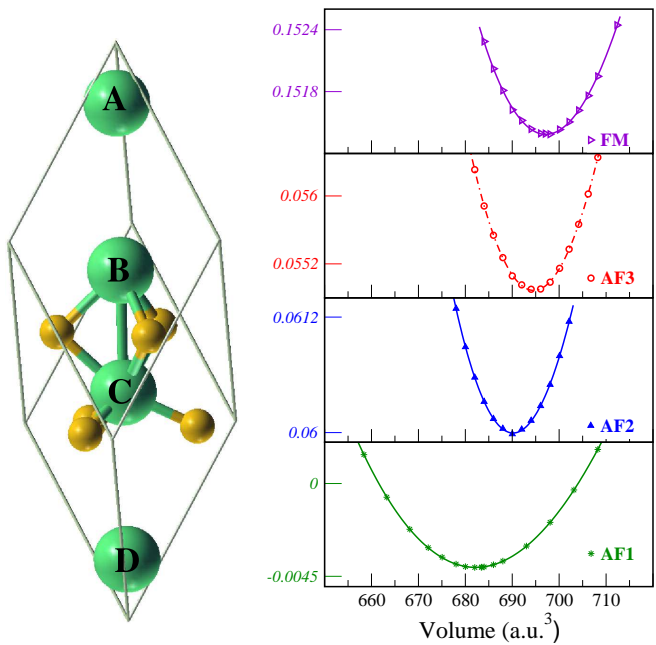}
\end{minipage}\hfill
\begin{minipage}[t]{0.60\linewidth}
\includegraphics[width=\linewidth,clip]{en-vol-all}
\end{minipage}\hfill
\caption{(color online) Left panel: The rhombohedral unit cell of
Fe$_2$O$_3$. The (big) green balls correspond to Fe atoms and brown ones
correspond to Oxygen atoms.  Right panel: Energy-Volume curves (symbols
correspond to calculated data and the curves are fits to data obtained
using Murnaghan's equation of state) for the four magnetic configurations
of hematite considered in this work. Note that the curve corresponding
to AF1 configuration (see text for notation) has the lowest energy at
the minimum and is therefore the ground state configuration.}
\label{en-vol-all}
\end{center}
\end{figure}

In the elemental rhombohedral cell, reported in Fig. \ref{en-vol-all},
four Fe atoms, spread along the z-axis, are grouped into two sets: two
Fe (B--C) atoms at a shorter distance, denoted by (Fe--Fe)$_{short}$
and a pair (A--B or C--D) at a larger distance, denoted by
(Fe--Fe)$_{long}$. Depending on the orientations of the magnetization of
the different Fe atoms, there can be four possible magnetic configurations
(AF1 $\equiv$ ABCD :: \(++--\), AF2 $\equiv$ ABCD :: \(+-+-\), AF3
$\equiv$ ABCD :: \(+--+\) and FM $\equiv$ ABCD :: \(++++\), where \(+\)
and \(-\) correspond to opposite magnetization directions). The optimized
lattice parameters as extracted from energy-volume curves are 10.277,
10.317, 10.340, 10.355 {\it a.u.} respectively for the AF1, AF2, AF3
and FM configurations. The most stable magnetic configuration is AF1,
as can be seen from the curves in Fig. \ref{en-vol-all}. These lattice
parameters were then used as the starting point for the subsequent cell
optimization at finite pressures. In all antiferromagnetic configurations
the Fe magnetic moments are close to 3.8 $\mu_{B}$/atom at 0 GPa. Moreover
all antiferromagnetic solutions are insulating while the ferromagnetic
solution is metallic.

\par
\begin{figure}[h]
\includegraphics[scale=0.42,clip]{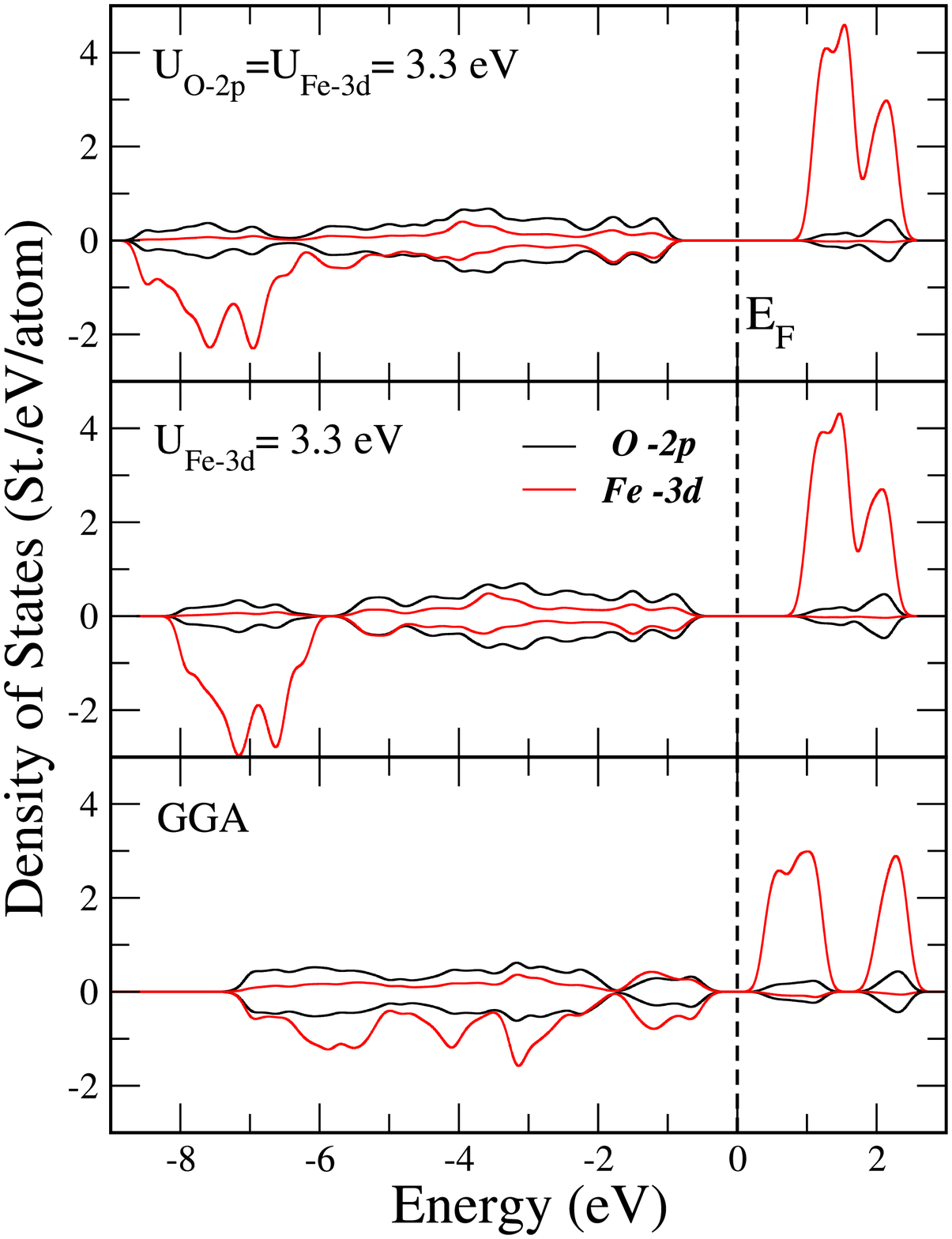}
\caption{(color online) Projected Density of States for the hematite (AF1) at 0 GPa per atom for GGA and GGA + $U$ calculations.}
\label{hem-dos}
\end{figure}

The partial density of states at 0 GPa, for AF1, is shown in the lowest
panel of Fig. \ref{hem-dos}. From the figure it is evident that the top
of the valence band is occupied by the 2$p$ states of O and the 3$d$
states of Fe on an equal basis. The valence and conduction band
edges are seperated by a well defined gap of about 0.3 eV. It is to be
noted that, all the antiferromagnetic configurations show a similar
valence and conduction edges, with similar value of the fundamental
gap, while the FM configuration is metallic. The experimental scenario
is a bit different: the top of the valence band has mainly O-2$p$ 
character, and the experimental value of the band gap is
2.0 eV [\onlinecite{Mochizuki-pss77}]. Both these deficiencies can be
attributed to an incomplete treatment of electronic correlation at the
LDA or GGA level. Similar results have been reported earlier and, for
instance, the calculated values of the band gap is 0.3 eV by employing
pseudopotential \cite{Rollmann-prb04}, 0.75 eV by augmented spherical
wave method \cite{Sandratskii-jpcm96} and 0.51 eV by linear muffin-tin
orbital \cite{Punkkinen-jpcm99}, respectively. The band gap value in
a Hartree Fock calculation \cite{Catti-prb95} is instead of grossly
overestimated as 11 eV due to the complete neglect of correlation effects.
A good representation of the band gap and of the atomic decomposition
of the band edge states can be obtained by including strong
correlation effects via $GGA+U$, as can be seen in the upper panels of
Fig. \ref{hem-dos}.  We wish to emphasize here, however, the qualitative
agreement of the GGA results with experiments of the basic electronic
structure of hematite.

\par
\begin{figure}[tbp]
\includegraphics[scale=0.29,clip]{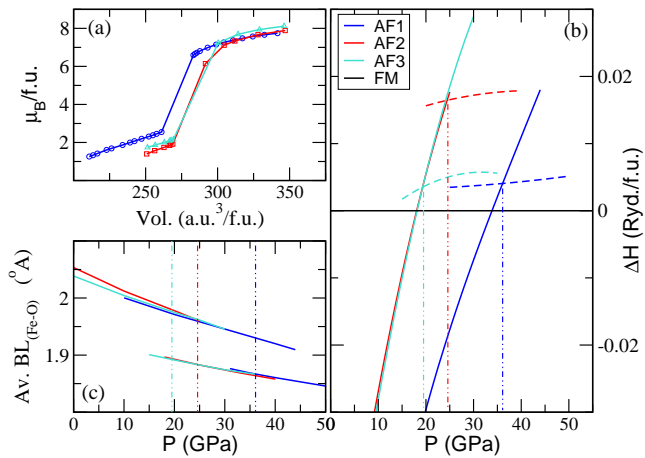}
\caption{(color online) Results for the different magnetic configurations in the hematite
structure. (a) Magnetic moment as a function of volume calculated at
the GGA level. (b) Relative enthalpy curves. Solid lines (vertical)
correspond to high spin (HS) and dashed lines (horizontal) are for low
(LS: 1$\mu_{B}$/Fe atom) spins. The reference enthalpy (base line)
corresponds to the FM configuration in the LS phase. (c) Dependence of
the average Fe-O bond-lenght (in the octahedra) with pressure. Vertical
dotted lines correspond to the spin transition pressures for different
magnetic configurations. For the FM configuration, not shown in figure,
similar characteristics are observed but at much smaller pressure of
around 5 GPa. }

\label{enthalpy-hem}
\end{figure}

To check the relative stability of the different magnetic
configurations af finite pressures we have calculated their enthalpies
by fully optimizing each of the structures by variable-cell damped
dynamics at several pressures.  The results are collected in
Figure.~\ref{enthalpy-hem}

\subsection{spin transition}
As can be seen in Fig. \ref{enthalpy-hem}(a), for all the magnetic
phases the value of the magnetic moment per Fe atom remains almost
constant at about 3.8 $\mu_{B}$/Fe atom at low pressure (large volumes).
With decreasing volume (increasing pressure) there is a sudden collapse
in magnetic moment magnitude to $\sim$ 1$\mu_{B}$/Fe atom.

The relative enthalpy curves of HS and LS phases for the various
configurations are depicted in Fig. \ref{enthalpy-hem}(b). The
crossover from the HS to LS occurs at around 36, 24.5 and 19.5 GPa,
for the AF1, AF2, AF3 configurations, respectively, and at $\sim$
5 GPa for the FM (not shown in the figure). From this figure it is
clear that the AF1 configuration is the most stable one upto a pressure
of about 34 GPa, where the FM phase becomes more stable.  This trend
was also observed by Rollman et.al..~\cite{Rollmann-prb04} Overall,
the computed sequence of relative stability of the different magnetic
phases of hematite is in agreement with experiment and previous
calculations.\cite{Rollmann-prb04,Sandratskii-jpcm96}

\begin{table*}[!tbp] \caption{\label{table1}Variation of Fe-Fe
bond lengths and of the average Fe-O bond lengths in the octahedral
coordination across the spin transition.  Increases ($\uparrow$) and
decreases($\downarrow$) in bond lengths are marked by arrows.  The values
of the distortion index, $\delta d_{Fe-O}$, and of the octahedral angle
variance, $\Delta\theta$, at the transition are also reported.}
\begin{ruledtabular} 
\begin{tabular}{cccccccccc}
 &&$(Fe-Fe)_{short}$&$(Fe-Fe)_{long}$&$(Fe-O)$&$\delta
 d_{Fe-O}$&$\Delta\theta^2$&\\
 &&$(\AA)\downarrow$&$(\AA)\uparrow$&$(\AA)\downarrow$&&$(\deg.^{2})$&\\
\hline Hematite:\\ 
\quad AF1 && 2.725 - 2.525 & 3.792 - 4.043 & 1.929 - 1.867 & 0.0427 - 0.0248 & 77.8 - 34.0 \\ 
\quad AF2 && 2.787 - 2.508 & 3.786 - 3.946 & 1.959 - 1.883 & 0.0298 - 0.0234 & 85.2 - 36.5 \\ 
\quad AF3 && 2.730 - 2.559 & 3.826 - 4.077 & 1.976 - 1.892 & 0.0480 - 0.0246 & 81.0 - 35.8 \\ 
\\ 
\end{tabular} 
\end{ruledtabular} 
\end{table*}

This HS-LS spin collapse at the transition is also accompanied by a sudden
change in the magnitude of a number of inter- and intra- octahedral
structural parameters. Notable inter-octahedral structural parameters
are the (Fe--Fe)$_{short}$ and (Fe--Fe)$_{long}$ distances, which
decrease and increase, respectively, at the spin transition as can be
seen from Table \ref{table1}. The same was observed by Rollman
et.al.~\cite{Rollmann-prb04} in their GGA calculations.

Although also the remote surroundings ({\it i.e.} the next to
next-nearest neighbors and inter-octahedral distances) have
non-negligible effects on the nature of the spin transition,
the most significant changes observed at the transition concern
mainly the octahedra, as already pointed out by a number of
studies.\cite{Takemoto-ic73,Gutlich-sb81,Konig-pic87,Konno-bcsj91,
Moliner-ica99,Guionneau-jmc99, Marchivie-ac03}
Of the intra-octahedral structural parameters, the average Fe-O bond
lengths, as shown in Fig. \ref{enthalpy-hem}(c), show a jump from higher
to lower values and display two very distinct and well separated slopes
in the HS and in the LS configurations.

Two other intra-octahedral structural parameters that displays significant
changes at the transition are the Fe--O bond-length distortion and
the octahedral-angle variance. Fe--O bond-length distortion, $\delta
d_{Fe-O}$, is defined as \(\delta d_{Fe-O} = \displaystyle \frac{1}{D}
\displaystyle\sum^{6}_{i=1} |\Delta d_{i}|\), where D is the average
(octahedral) Fe--O bond-length at that particular volume and $\Delta
d_{i}$ is the difference of the {\it i-th} Fe--O bond-length from its
average. The octahedral-angle variance\cite{Robinson-science71} is instead
given by \(\Delta \theta^2 = \displaystyle\sum^{12}_{j=1} (\theta_{j}
- 90)^{2}/11\). Where \(\theta_{j}\) is the variation of the {\it j-th}
O--Fe--O angle from its ideal value of 90$^{0}$.

At the spin transition, both the mentioned intra-octahedral
structural parameters, $\delta d_{Fe-O}$ and $\Delta \theta^2$,
show a jump from a larger to a smaller value as can be seen
from Table. \ref{table1}. This indicates that at the spin transition
the octahedra become more regular and this happens irrespective of the
octahedral surroundings.\cite{note-on-transition-parameters} Moreover,
the sudden jump of these parameters can directly be related with the
$\sim$10\% volume collapse that is observed for all the magnetic phases
at the spin transition.

\subsection{structural transition}
Along with the spin transition discussed above, $\alpha$-Fe$_{2}$O$_{3}$
also undergoes structural phase transition/s under pressure and
temperature. Two candidate structures for this phase are the
orthorhombic perovskite ({\it opv}) with {\it Pbnm} symmetry and the
Rh$_{2}$O$_{3}$(II) with {\it Pbcn} symmetry.

\par
\begin{figure}[h]
\includegraphics[scale=0.52,clip]{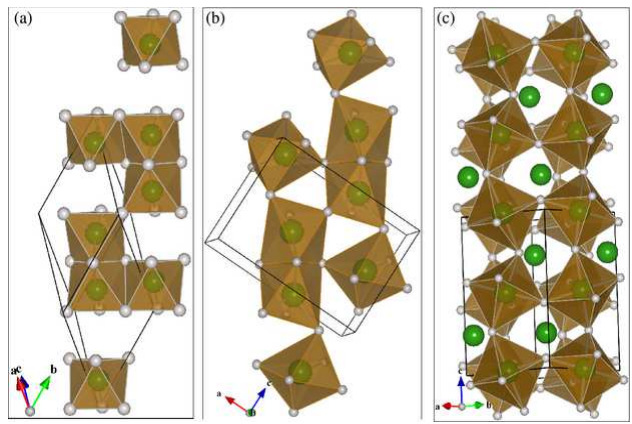}
\caption{(color online) Polyhedral structures of Fe$_{2}$O$_{3}$ in different crystal packing. (a): hematite, (b): Rh$_{2}$O$_{3}$(II), (c): Pbnm type perovskite. Small white spheres are oxygen atoms and big green ones are for the Fe atoms.}
\label{struc-hem-pbcn-pbnm}
\end{figure}

In hematite ($\alpha$-Fe$_{2}$O$_{3}$), in the corundum structure,
the Fe atoms sit inside the cage of a single kind of polyhedra which
are octahedra. Each of these octahedra shares three edges and one face
with the neighbouring octahedra. Rh$_{2}$O$_{3}$(II) structure with {\it
Pbcn} symmetry is of the distorted corundum type and also has the same
single type octahedral arrangement. Owing to a larger octahedral distortion
and rotation of the octahedra in Rh$_{2}$O$_{3}$(II), one of the three
sharing edges is broken by the alteration of one of the coordinating
oxygens. This crystallographic change is minor as can be seen from the
Fig. \ref{struc-hem-pbcn-pbnm}: instead of three shared edges as in
hematite, octahedra in Rh$_{2}$O$_{3}$(II) have two sharing edges with
the addition of a corner sharing bonding.

On the other hand, for the pervoskite structure with {\it Pbnm}
symmetry, there are two types of polyhedral environment surrounding
Fe atoms. The six-fold coordinated (octahedra) and the 8-fold
coordinated polyhedra are arranged in alternate layers along the $z$
direction. The octahedral connectivity is via corner sharing, while the
8-fold coordinated polyhedra are connected to each other by edge and
corner sharing. The interpolyhedral connectivity (between octahedra and
8-fold coordinated polyhedra) is maintained via face and edge sharing.

\par
A M\"ossbauer spectroscopy study at 300K by Pasternak {\it
et. al.}\cite{Pasternak-prl99} showed that the transition from hematite to
the high-pressure phase starts at about 45 GPa and completes at about 72
GPa. The existence of {\it opv} phase at high-pressure was ruled out by
pointing out the single cationic nature of the Fe-ions. Moreover, Fe-ions
were characterized as non-magnetic and the magnetic to non-magnetic
transition connected to the structural phase transition. In a later study
Badro {\it et. al.}\cite{Badro-prl02} could separate the crystallographic
transition from the spin transition. According to this study the spin
transition is isostructural and takes place in the high-pressure phase. In
a more recent study Ono {\it et. al.}\cite{Ono-jpcs04,Ono-jpcm05}
reported the transition pressure to be 26.2 GPa at 300 K. It was
concluded there that high-temperature heating played a fundamental
role in overcoming the kinetic barrier hindering the phase transition,
thus lowering the transition pressure. The combined experimental and
theoretical study by Rozenberg {\it et. al.}\cite{Rozenberg-prb02}
also indicated Rh$_{2}$O$_{3}$(II) as the HP2 phase. The {\it opv}
phase was found to be higher in energy than the Rh$_{2}$O$_{3}$(II)
phase by as much as 25 Ryd/f.u.\ at 76 GPa. However, no other detailed
theoretical study exist on the relative stability of these two phases.

\par
\begin{figure}[h] \includegraphics[scale=0.30,clip]{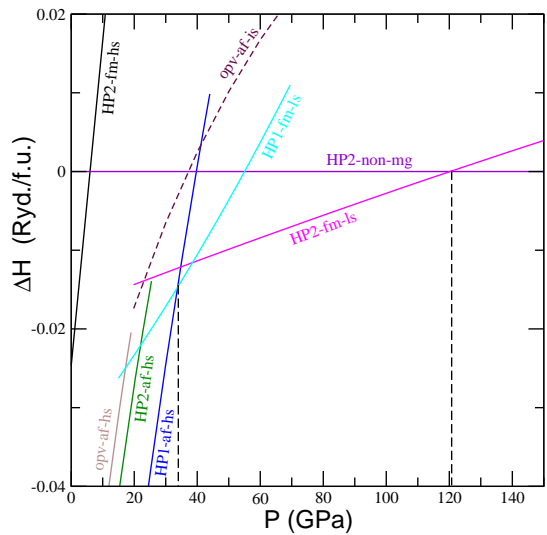}
\caption{(color online) Relative enthalpy curves for the Rh$_{2}$O$_{3}$(II)
(HP2), {\bf \emph{opv}} (opv) along with the corundum structure
(HP1). HP1 structure in the LS phase is dynamically stable upto about
180 GPa. For the opv structure in the FM configuration, not shown here,
the relative enthalpy curve remain well above the base line here. The
dashed vertical lines are guide to the eye to locate the transition
pressures.} \label{enthalpy-wrt-pbcn-nonmg} \end{figure}

We have calculated the static enthalpies for these two structures,
both in FM and AF configurations, for a finite range of pressures.
Fig.~\ref{enthalpy-wrt-pbcn-nonmg} shows the relative enthalpy-pressure
curves for these two candidate structures along with the AF1 corundum
structure. As can be seen from the figure, at low pressures all structures
show an HS solution. Corundum in the HS AF1 configuration is the most
stable one upto about 34 GPa. At this pressure the enthalpy of corundum
in the LS FM configuration becomes lower. This crossover from HS AF to
LS FM is also observed for Rh$_{2}$O$_{3}$(II) structure at about 26
GPa (as can be seen from the crossing of the {\it HP2-af-hs} and {\it
HP2-fm-ls curves} in Fig.~\ref{enthalpy-wrt-pbcn-nonmg}). The crossover
from HS to LS is also present for {\it opv} structure, but only in the
octahedral sites. The 8-fold coordinated polyhedra remain in the high
spin and hence, overall, the system remain in a kind of intermediate
mixed (half HS and half LS) spin state. We mention also that from
$\sim$ 34 GPa to about 38 GPa the LS FM configuration in the corundum
structure is the most stable one. After this, in the broad range of
pressure from $\sim$ 38 GPa to almost 120 GPa the Rh$_{2}$O$_{3}$(II)
structure in the LS FM configuration is the moste stable. With increasing
presssure there is a gradual decrease of magnetization in the system
that finally becomes non-magnetic from $\sim$ 120 GPa onwards. Hence,
upto the pressure range shown in this figure, {\it opv} does not have
any stability region. Therefore, our calculations, in agreement with recent
experiments,\cite{Pasternak-prl99, Badro-prl02, Rollmann-prb04} rule
out the two cationic site picture, {\it i.e.} {\it opv}, and establishes
Rh$_{2}$O$_{3}$(II) as the HP2 or next stable phase after the corundum
one..

\subsection{ further structural transitions}
\par
The high temperature experiments by Ono {\it et. al.}\cite{Ono-jpcs04}
reported of another structural phase transition after the HP2
phase. Their data show that the transformation of this phase begins
at around 60 GPa and 2500K. They identified this phase as CaIrO$_{3}$
type with Cmcm symmetry. However, this designation was made without a
detailed structural analysis in analogy to other systems. The prototype
sesquioxide, Al$_{2}$O$_{3}$, for instance, having corundum structure
at ambient conditions, undergoes a structural transformation from
Rh$_{2}$O$_{3}$(II) type phase to a post-perovskite one of CaIrO$_{3}$
type, and this structure has been found as the post-perovskite phase in
most of the instances of post-perovskite phases discovered till today
[see Ref.\onlinecite{Tsuchiya-prb05} and references therein].

\begin{figure}[h]
\includegraphics[scale=0.59,clip]{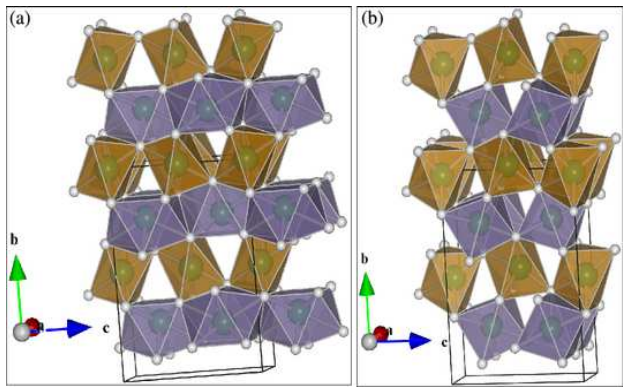}
\caption{(color online) Two candidate high pressure structures of Fe$_{2}$O$_{3}$. (a)
HP3: Cmcm type post-perovskite and (b) HP4: Pmc2$_{1}$ type
post-perovskite. Small white spheres are oxygen atoms and big green ones
are for the Fe atoms.}
\label{struc-cmcm-pmc21}
\end{figure}
 
\par
In Fe$_{2}$O$_{3}$, being another sesquioxide, it is therefore natural to
explore the possible higher pressure phases starting with CaIrO$_{3}$
structure with Cmcm symmetry (denoted as HP3 from here). In terms of
polyhedral arrangements this structure has two types of polyhedral
arrangements, {\it i.e.} octahedra and 8-fold coordinated polyhedra.
The difference with respect to the Pbnm type structure is that there
is no longer any inter-layer octahedral connection. Instead alternate
layers of octahedra and 8-fold coordinated polyhedra form the structure
as shown in Fig. \ref{struc-cmcm-pmc21}(a).

\begin{figure}[h]
\includegraphics[scale=0.3,clip]{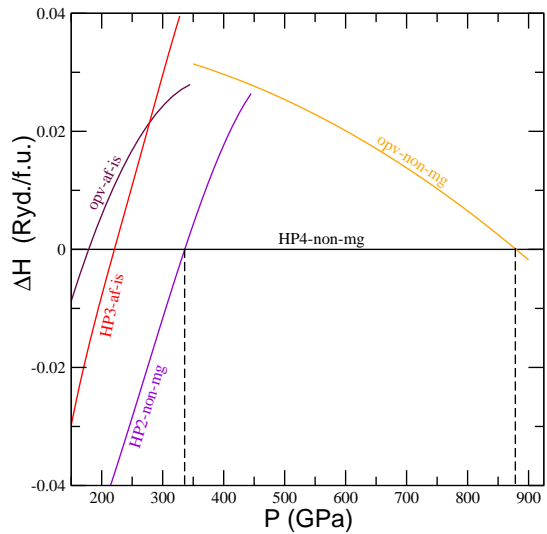}
\caption{(color online) Relative enthalpy curves for the Rh$_{2}$O$_{3}$(II) (HP2),
{\bf \emph{opv}} (opv), Cmcm type post-perovskite (HP3) and Pmc2$_{1}$
type post-perovskite (HP4). The dashed vertical lines are guide to the
eye to locate the transition pressures.}
\label{enthalpy-wrt-pmcn-nonmg}
\end{figure}

In Fig. \ref{enthalpy-wrt-pmcn-nonmg}, one can see that around 200 GPa
both {\it opv} and the HP3 phases are in the intermediate spin state
(Fe in the octahedral cage is in low spin state, while it is in a
high-spin state when 8-fold coordinated to Oxygen) and are dynamically
stable. However, upto about 337 GPa, the enthalpy of the nonmagnetic HP2
phase is lower. From 337 GPa onwards, the enthalpy of the nonmagnetic
phase of a new structure\cite{note1} crosses the HP2 one. This new
structure, denoted HP4 in the following, has Pmc2$_{1}$ symmetry
with an orthorhombic unit cell consisting of 20 atoms. Pmc2$_{1}$
is a subgroup of Cmcm. The 4c,4a and 8f Wyckoff positions in Cmcm are
split into 2a and 2b Wyckoff positions in Pmc2$_{1}$ space group. This
splitting of the symmetry positions leads to different kind of polyhedral
arrangement in the new phase. Instead of alternate layers of octahedra
and of 8-fold coordinated polyhedra as in Cmcm, the arrangement in
Pmc2$_{1}$ corresponds to alternate layers of octahedra and of 7-fold
coordinated polyhedra. Compariing HP4 with HP3, it can be seen from
Fig. \ref{enthalpy-wrt-pmcn-nonmg} that the AF HP3 curve, labelled
{\it HP3-af-is}, crosses the non-magnetic HP4 curve ({\it HP4-non-mg})
at $\sim$ 225 GPa. This newly achieved structural phase (HP4) remains
stable for a wide pressure range. At much higher pressures, however,
the enthalpy of the nonmagnetic {\it opv} phase becomes competitive and
finally becomes the stable structure at about 880 GPa.

Experimentally,\cite{Ono-jpcs04} a post-perovskite type phase was obtained
at pressure (~60 GPa at 2500K) much below the one achieved here by our
calculations, where the post-perovskite phase is stable with respect to
the Rh$_{2}$O$_{3}$(II) phase only beyond 330 GPa.  High temperature
heating likely plays a role in reducing the experimental transition
pressure. For instance, for the structural transition from corundum-type
to Rh$_{2}$O$_{3}$(II), the transition pressure drops from 46 GPa to
26 GPa when the samples are laser heated from room temperature to about
2500K.\cite{Ono-jpcs04, Ono-jpcm05, Badro-prl02} We notice that the Cmcm
type post-perovskite structure proposed (without structural determination)
by Ono. ${\it et. al.}$\cite{Ono-jpcs04, Ono-jpcm05} is never the stable
one in our calculations, rather Pmc2$_{1}$ symmetry--a subgroup of
Cmcm--is found. Further efforts are needed to establish experimentally
the structural parameters of this post-perovskite phase.

This extreme condition phase might have implications in exploring and
understanding the rocky cores of solar giants and/or rocky mantles
of recently discovered terrestrial exoplanets.\cite{Guillot-pt04,
Rivera-aj05, Valencia-icarus06}.

\begin{figure*}[tbp]
\includegraphics[scale=0.45,clip]{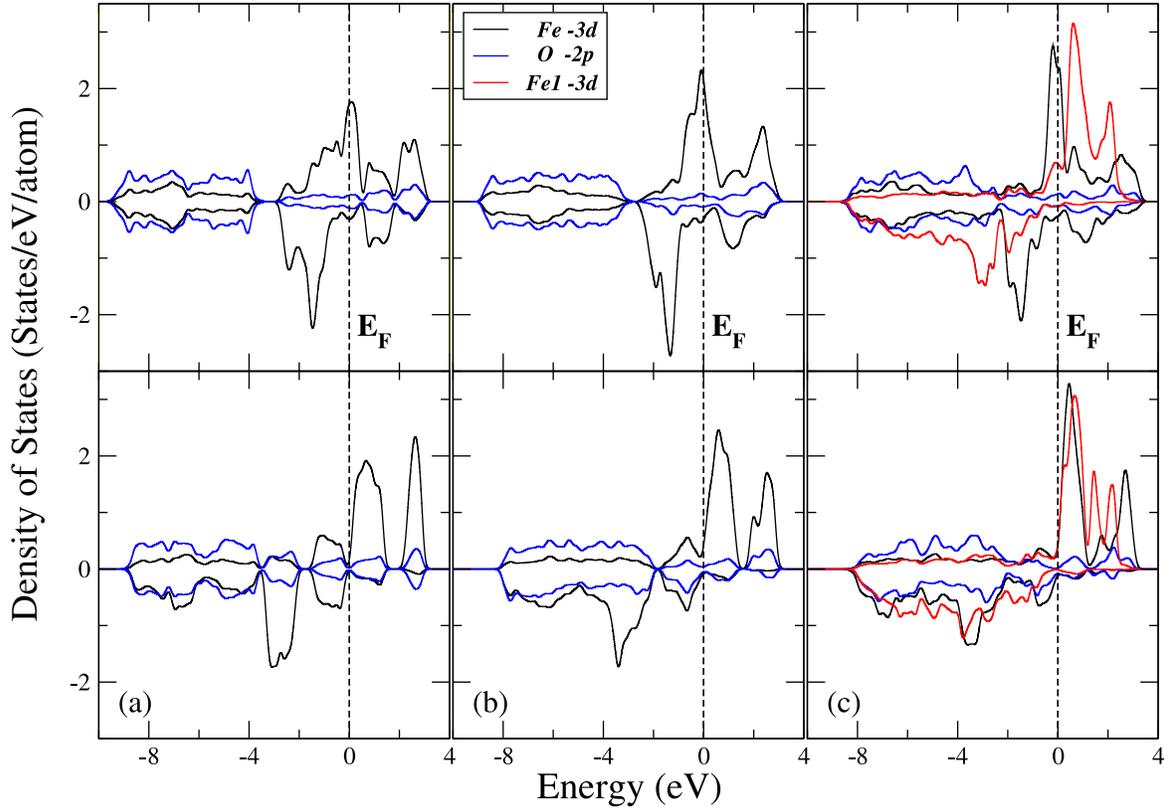}
\caption{(color online) Projected density of states in the vicinity of the
spin transition. Lower panels show the DOS just before the spin
transition occurs and the upper panels are the same just after
the spin transition. (a) for the hematite structure. (b) for the
Rh$_{2}$O$_{3}$(II) type Fe$_{2}$O$_{3}$. (c) for the Pbnm type
Fe$_{2}$O$_{3}$. Note that in Pbnm type Fe$_{2}$O$_{3}$ Fe is in two
types of polyhedral coordination. The spin transition occurs only in the
octahedral cage. Whereas the Fe in 8-fold coordinated polyhedra remain
in high spin.}
\label{dos-at-transition}
\end{figure*}

\begin{figure*}[tbp]
\includegraphics[scale=1.05,clip]{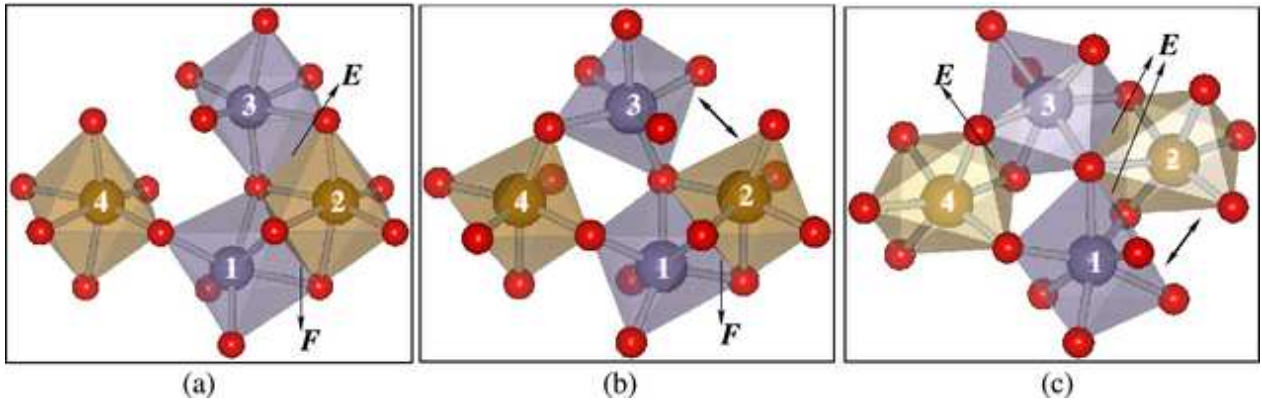}
\caption{(color online) Figure shows the evolution of the basic network of the
polyhedra. {\bf E} and {\bf F} correspond respectively to edge sharing
and face sharing bonding. Red (smaller) spheres are O atoms while
the bigger spheres at the centres of the polyhedrons correspond to Fe
atoms and are denoted by {\bf 1, 2, 3, 4}. (a) correspond to Hematite,
(b) Rh$_{2}$O$_{3}$(II) structure, and (c) Pmc2$_{1}$ structure. Note
the changes in the edge, face and corner sharing, leading to different
types of structures.}
\label{all-octa-str}
\end{figure*}

\subsection{in the vicinity of spin transition and structural evolution}
Before coming to conclusions we want to summarize two issues: What happens
at the spin transition and what is the overall structural evolution in
Fe$_{2}$O$_{3}$. We have already mentioned that the phenomena of spin 
transition occurs in the octahedral cages\cite{Ghosh1} for all the three 
phases that are dynamically
stable at ambient pressure: {\it i.e.} hematite, Rh$_{2}$O$_{3}$(II)
type Fe$_{2}$O$_{3}$ and Pbnm type Fe$_{2}$O$_{3}$. A vivid look into the
density of states profile, depicted in Fig. \ref{dos-at-transition}, shows
that at the transition, the $d$--band width increases substantially. Our
explanation for this goes as: with increasing pressure the O atoms gets
closer to the Fe atoms. This induces enhancement in the octahedral
crystal field splitting of the triply degenerate $t_{2g}$ and doubly
degenerate $e_{g}$ levels. At some point in the pressure scale the
strength of the crystal field energy overcomes the exchange energy
and the Fe atoms transit from a state with $^{3}t_{2g}$ $^{2}e_{g}$
configuration to a state with $^{5}t_{2g}$ $^{0}e_{g}$. Also, it can be
seen from Fig. \ref{enthalpy-wrt-pbcn-nonmg} that in terms of magnetic
state the system exists in three distinct regions, {\it i.e.}, with HS,
LS and nonmagnetic. The transition from LS to the nonmagnetic state
is rather gradual. This happens only when the hybridization strength
between Fe--$3d$ and O--$2p$ levels overcomes the crystal field energy.

To explain how the structure evolves, we present the sequence of basic
polyhedral network from hematite to Pmc2$_{1}$ type structure through
Rh$_{2}$O$_{3}$(II) structure in Fig. \ref{all-octa-str}. The face
sharing network in between ocathedron {\bf 1} and {\bf 2} in hematite
is retained in Rh$_{2}$O$_{3}$(II). However, the edge sharing network
between octahedron {\bf 2} and {\bf 3} of hematite is broken and instead
a corner sharing bonding is formed between octahedron {\bf 3} and {\bf 4}
in Rh$_{2}$O$_{3}$(II). In the transition from Rh$_{2}$O$_{3}$(II) to the
Pmc2$_{1}$ structure, the face sharing network between octahedron {\bf 1}
and {\bf 2} of Rh$_{2}$O$_{3}$(II) is broken and the rearrangement of
the internal parameters leads to two kinds of polyhedral connectivity,
one with 6-fold coordination ({\bf 2} and {\bf 4}) and the other with
7-fold coordination ({\bf 1} and {\bf 3}).

\section{Conclusion}
In this work, systematic density functional calculations
of Fe$_{2}$O$_{3}$ have been performed over a wide range of
pressure. Corundum type phase displays high-spin to low-spin transition
for all magnetic configurations. This corundum type phase transforms to
a distorted corundum or Rh$_{2}$O$_{3}$(II) type phase at about 38 GPa
while Pbnm type perovskite structure don't have any stability field in
this pressure regime. The phenomenon of spin transition is also present in
Rh$_{2}$O$_{3}$(II) structure. On increasing pressure, our calculations
identify a new structure with Pmc2$_{1}$ symmetry, a subgroup of Cmcm,
as the post-perovskite structure, while the experimentally proposed Cmcm
phase is never the most stable one. Our calculations also show that the
orthorhombic perovskite type structure with Pbnm symmetry might have a
stability field at extremely high pressures.

\acknowledgements

Calculations were performed at SISSA and at CINECA computing center in
Bologna, also thanks to INFM computing grants. All calculations have
been done employing the PWscf code in the Quantum Espresso\cite{QE}
distribution. For the graphics part VESTA\cite{Momma-ccc06} package has
been used in addition to the XCrySDen\cite{xcrysden} package.

\clearpage


\begin{thebibliography}{99}
\bibitem{note-on-transition-parameters}
Electronic spin transition is also observed in Pbcn and Pbnm type
Fe$_2$O$_3$. Pbnm type Fe$_2$O$_3$ has two types of polyhedral (octahedra
and 8-fold coordinated polyhedra) arrangement and the electronic spin
transition occurs only in the octahedra. At the spin transition
the octahedral parameters jump from a higher to a lower value. See
[\onlinecite{Ghosh1}] for details.
\end{thebibliography}
\end{document}